\documentclass[pr, 12pt, superscriptaddress, showpacs]{revtex4}  
\usepackage{amssymb}
\usepackage{amsmath}
\usepackage{bbm}
\usepackage{graphicx}
\usepackage{color}

\begin{document}

\title{Ultrafast switching of composite order in A$_3$C$_{60}$}

\author{Philipp Werner}
\affiliation{Department of Physics, University of Fribourg, 1700 Fribourg, Switzerland}
\author{Hugo Strand}
\affiliation{Department of Physics, University of Fribourg, 1700 Fribourg, Switzerland}
\author{Shintaro Hoshino}
\affiliation{RIKEN Center for Emergent Matter Science (CEMS), Wako, 351-0198 Saitama, Japan}
\author{Martin Eckstein}
\affiliation{Max Planck Research Department for Structural Dynamics, University of Hamburg-CFEL, 22761 Hamburg, Germany}


\begin{abstract}
We study the controlled manipulation of the Jahn-Teller metal state of fulleride compounds using nonequilibrium dynamical mean field theory. This anomalous metallic state is a spontaneous orbital-selective Mott phase, which is characterized by one metallic and two insulating orbitals. Using protocols based on transiently reduced hopping amplitudes or periodic electric fields, we demonstrate the possibility to switch orbitals between Mott insulating and metallic on a sub-picosecond timescale, and to rotate the order parameter between three equivalent states that can be distinguished by their anisotropic conductance. The  Jahn-Teller metal phase of alkali-doped fullerides thus provides a promising platform for ultrafast persistent memory devices. 
\end{abstract}

\date{\today}

\maketitle

\hyphenation{}

\section{Introduction}

The manipulation and switching of ordered states by optical pulses is an active research field which connects fundamental theoretical issues to important technological applications. Two decades ago Beaurepaire {\it et al.} showed that short laser pulses can melt the ferromagnetic spin order in Ni in less than a picosecond \cite{Beaurepaire1996}. In ferrimagnets, such as Gd-Fe alloys, the energy injected by light pulses can result in a switching of the sublattice magnetization \cite{Radu2011,Chimata2015}. More recently, it was shown that the polaronic Mott insulator TaS$_2$ can be switched from the insulating commensurate charge density wave phase into a conducting so-called ``hidden state" (with an unconventional form of charge order) by applying short laser pulses \cite{Stojchevska2014}. 

The investigation of such light-induced switching mechanisms is at least partially motivated by the quest for a fast and energy-efficient memory technology \cite{Vaskivskyi2016}. Over the last decades the performance of memory devices has lagged behind the rapid gains in processing power. Desirable features of a high-performance memory include electrically or optically controlled fast write and erase operations, fast read-out, low energy consumption and a non-volatile nature of the stored information. Ideally, the free energy landscape is characterized by at least two minima corresponding to physically distinct realizations of the order, and the switching between these minima can be accomplished without a transient melting of the order, through the application of a suitable force which rotates the order parameter. If this rotation furthermore involves no displacement of charge, and does not couple to static lattice distortions, it may be potentially accomplished on a very fast, electronic timescale, and at little energy cost.

While existing memory technologies are based on ``conventional" order parameters such as magnetization or charge order, we will explore in this work a fundamentally different, and rather exotic type of ordered state, which involves a symmetry breaking on the two-particle level (``composite order"). We will show that this type of order, which has recently been detected in Rb$_x$Cs$_{3-x}$C$_{60}$ \cite{Zadik2015}, and theoretically explained in Ref.~\onlinecite{Hoshino2016}, encompasses a range of desirable properties which make it interesting for memory applications.

Alkali-doped fulleride compounds have been investigated extensively because of their unconventional form of high-temperature superconductivity. These systems can be described by a half-filled three orbital Hubbard model with degenerate bands derived from the t$_{1u}$ molecular orbitals. The characteristic and unusual feature of this class of materials is an effectively negative Hund coupling, resulting from the overscreening of the small bare Hund coupling by Jahn-Teller phonons \cite{Fabrizio1997,Capone2009}. Such three-orbital systems with negative Hund coupling exhibit, in an experimentally relevant range of interactions or pressure, a spontaneous orbital selective Mott transition \cite{Hoshino2016}. This is a transition into a long-range ordered phase with a composite order parameter, namely an orbital-dependent double occupation. While different types of such orders exist, the most stable one features one metallic and two insulating orbitals. There are thus three different realizations of this most stable order, which can be distinguished by their conductance anisotropy, and which may be used to store information. Since the spontaneous orbital-selective Mott (SOSM) phase has no ordinary orbital moment, i.e. all three orbitals remain half-filled, this electronic order should not induce static lattice distortions and it may thus be possible to switch it efficiently between the different physical realizations of the order parameter. Alkali-doped fullerides in the SOSM phase hence provide a potential platform for ultra-fast and energy efficient persistent memory devices.

\section{Model and method}

The essential physics of alkali doped fullerides is captured by the half-filled three orbital Hubbard model with negative Hund coupling \cite{Capone2009}. We consider the Hamiltonian
\begin{equation}
H_\text{latt}(t) = -\sum_{ i,j, \alpha,\sigma} v_{ji,\alpha}(t) d^\dagger_{j,\alpha,\sigma}d_{i,\alpha,\sigma} -\mu\sum_{i,\alpha,\sigma} n_{i,\alpha,\sigma}+H_{\text{int},i}, 
\end{equation}
with $d_{i,\alpha,\sigma}$ the annihilation operator for an electron with spin $\sigma$ in orbital $\alpha$ at site $i$, $\mu$ the chemical potential, and $n_{\alpha,\sigma}=d^\dagger_{\alpha,\sigma} d_{\alpha,\sigma}$. The orbitals are labeled according to their symmetry $\alpha=x,y,z$. Assuming for simplicity nearest-neighbor hopping on a cubic lattice, the symmetry of the orbitals implies that hopping occurs only between orbitals of the same type, with $v_{ji,\alpha}\equiv v$ if the nearest-neighbor bond $(ji)$ is directed along $\alpha$, and $v_{ji,\alpha}=0$ otherwise. 
Note that this orbital-dependent hopping implies an anisotropic conductance of the SOSM phase.
The interaction term $H_{\text{int},i}$ is of the Slater-Kanamori form,
\begin{align}
H_\text{int} =& \sum_\alpha Un_{\alpha,\uparrow} n_{\alpha,\downarrow}+\sum_{\alpha\ne\beta} (U-2J)n_{\alpha,\uparrow}n_{\beta,\downarrow}+\sum_{\alpha>\beta,\sigma}(U-3J)n_{\alpha,\sigma}n_{\beta,\sigma}\nonumber\\
&+\gamma(H_\text{spinflip}+H_\text{pairhop}). 
\label{Hint}
\end{align}
and we choose $J/U=-1/4$. This value of $|J/U|$ is substantially larger than in  realistic compounds \cite{Nomura2015}, where it is around $0.02$-$0.03$, but the physics does not depend qualitatively on the value of $J$ and the above choice is advantageous from a numerical point of view \cite{Hoshino2016}. Furthermore, it was shown that while the pair-hopping term stabilizes the SOSM phase, the composite order also appears in the Ising approximation ($\gamma=0$), so we will first discuss the results for this simpler model. 

To describe the laser excitation which is later used to switch between different SOSM phases we include an electric field in the model, by multiplying the hopping $v_{ji,\alpha}$ with a time-dependent Peierls factor $e^{i\phi_{ji}(t)}$, where the phase $\phi_{ji}(t)=e\vec{A}(t)(\vec{r}_j-\vec{r}_i)$ is the projection of the vector potential $\vec{A}(t)$ onto the bond vector $(\vec{r}_j-\vec{r}_i)$, and the vector potential $\vec{A}(t)$ is related to the electric field by $\vec{E}(t)=-d\vec{A}(t)/dt$. Due to the orbital-dependent anisotropy of the hoppings, one can therefore selectively modify the hoppings for the three orbitals by a suitable choice of the time-dependent polarization of the field, which determines the components
$A_\alpha$ along the $\alpha$-direction.

To solve the time-dependent lattice problem, we use the nonequilibrium dynamical mean-field theory \cite{Aoki2014}, which maps the lattice system to an auxiliary three-orbital impurity problem with the same interactions as Eq.~(\ref{Hint}), subject to a self-consistency condition for the hybridization function. For the orbital-dependent hoppings introduced above, we obtain the hybridization function at an arbitrary lattice site $j$ from the cavity method \cite{Georges1996},
\begin{equation}
\Delta_\alpha(t,t')=
\sum_{l} v_{jl,\alpha}(t) G_{l,\alpha}^{[j]}(t,t') v_{lj,\alpha}(t').
\end{equation}
Here $G_{l,\alpha}^{[j]}(t,t')$ is the cavity Green's function, i.e., the Green's function on site $l$ for a system with site $j$ removed from the lattice. 
With two nearest neighbors for each orbital we have 
\begin{align}
\Delta_\alpha(t,t')&=v(t)e^{i\phi_\alpha(t)}G_{\alpha}(t,t')v(t')e^{-i\phi_\alpha(t')}+v(t)e^{-i\phi_\alpha(t)}G_\alpha(t,t')v(t')e^{i\phi_\alpha(t')}\nonumber\\
&=2v(t)\cos(\phi_\alpha(t))G_\alpha(t,t')v(t')\cos(\phi_\alpha(t'))+2v(t)\sin(\phi_\alpha(t))G_\alpha(t,t')v(t')\sin(\phi_\alpha(t')),
\label{eq_field}
\end{align}
where $\phi_\alpha(t) = eaA_\alpha(t)$, with the lattice spacing $a$. For simplicity we have also replaced $G_{l,\alpha}^{[j]}(t,t')$ by $G_{\alpha}(t,t')$, which  approximates the orbital-dependent local density of states by a semi-ellipse of bandwidth $W=4\sqrt{2}v$
corresponding to an infinitely connected Bethe lattice. 
We expect that the shape of the density of states does not qualitatively affect the results presented below. We choose $W=0.4$ eV in the following, to set this energy scale to a value approximately equal to the bandwidth of A$_3$C$_{60}$ \cite{Nomura2012}.

The nonequilibrium impurity model is treated with a strong-coupling perturbative impurity solver \cite{Eckstein2010}. The latter approach is suitable for the strongly correlated regime that we are interested in, and it can be straightforwardly applied to multiorbital problems by introducing pseudo-particles for each atomic eigenstate \cite{Keiter1971}. To enable an efficient simulation, we have automatized the detection of degeneracies in the large set of self-energy and Green function diagrams, which allows us to reach the relevant timescales both at the first order (non-crossing approximation, NCA) and second order (one-crossing approximation, OCA) level. Specifically, for the half-filled model with $\gamma=0$, the symmetry analysis reduces the number of NCA self-energy diagrams from 384 to 156, and the number of NCA Green function diagrams from 192 to 96.

To mimic energy dissipation to the lattice and to explore the heating effect on the order parameter dynamics we locally couple a bosonic heat bath to the electrons \cite{Eckstein2013}. This is done by adding a self-energy of the form 
\begin{equation}
\Sigma^\text{bath}_\alpha(t,t')=g D_0^{(\omega_0,\beta)}(t,t')G_\alpha(t,t')g
\end{equation}
to the hybridization function of each orbital, where $D_0$ is the equilibrium propagator of a boson with energy $\omega_0$ at inverse temperature $\beta$ and $g$ is the electron-boson coupling strength \cite{Aoki2014}. 

\begin{figure}[t]
\includegraphics[angle=-90, width=0.32\columnwidth]{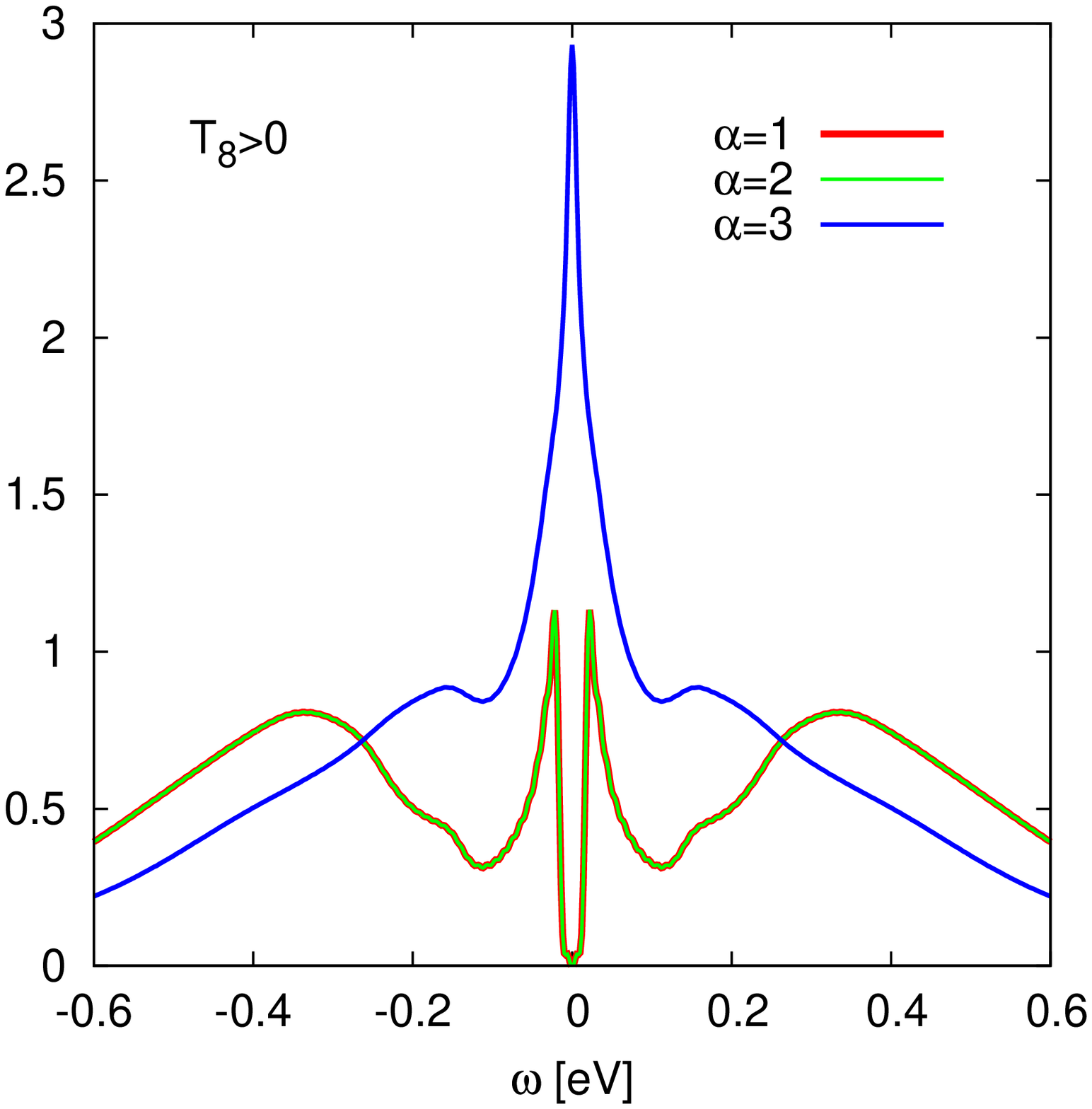}\hfill
\includegraphics[angle=-90, width=0.32\columnwidth]{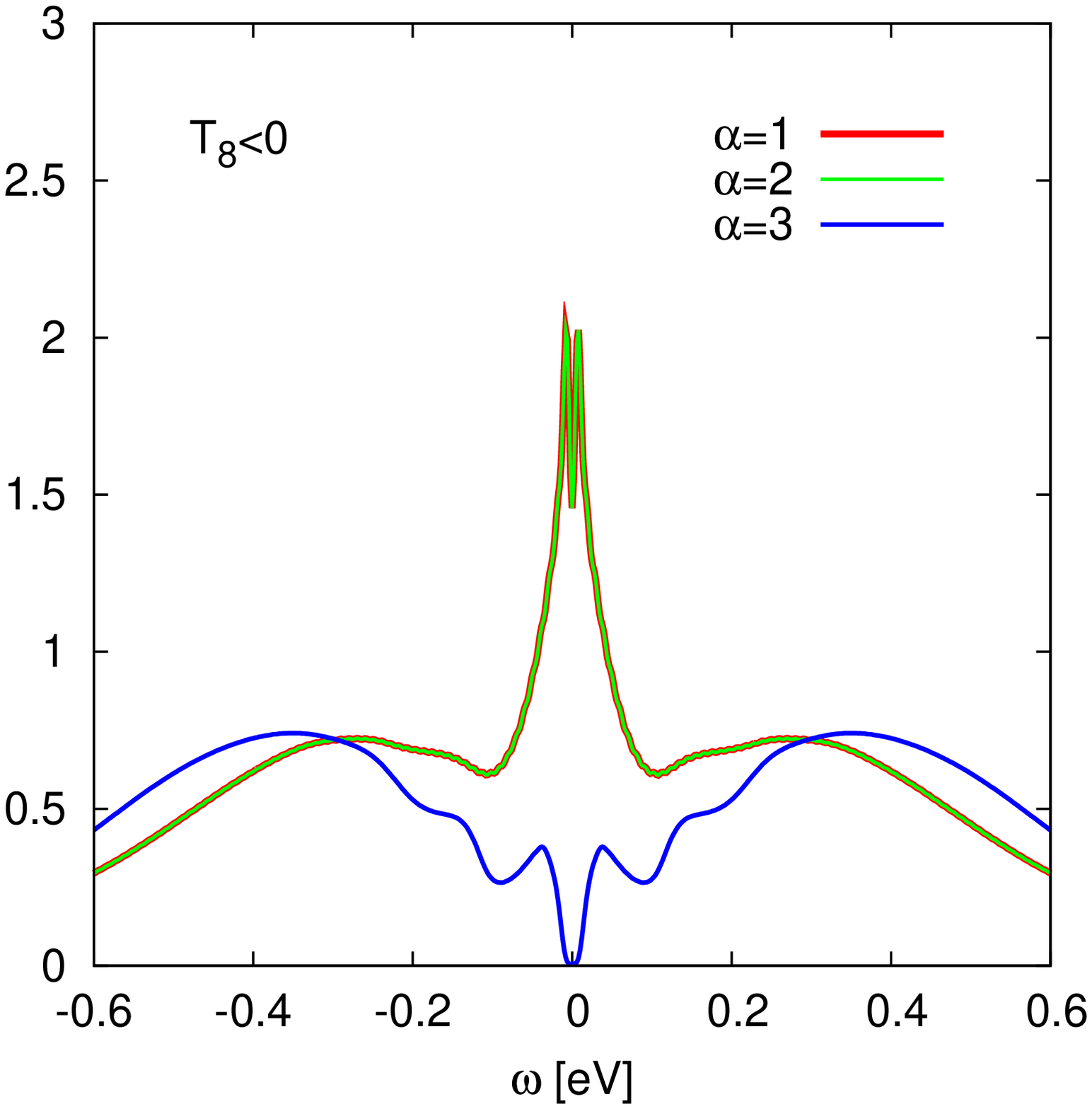}\hfill
\includegraphics[angle=-90, width=0.29\columnwidth]{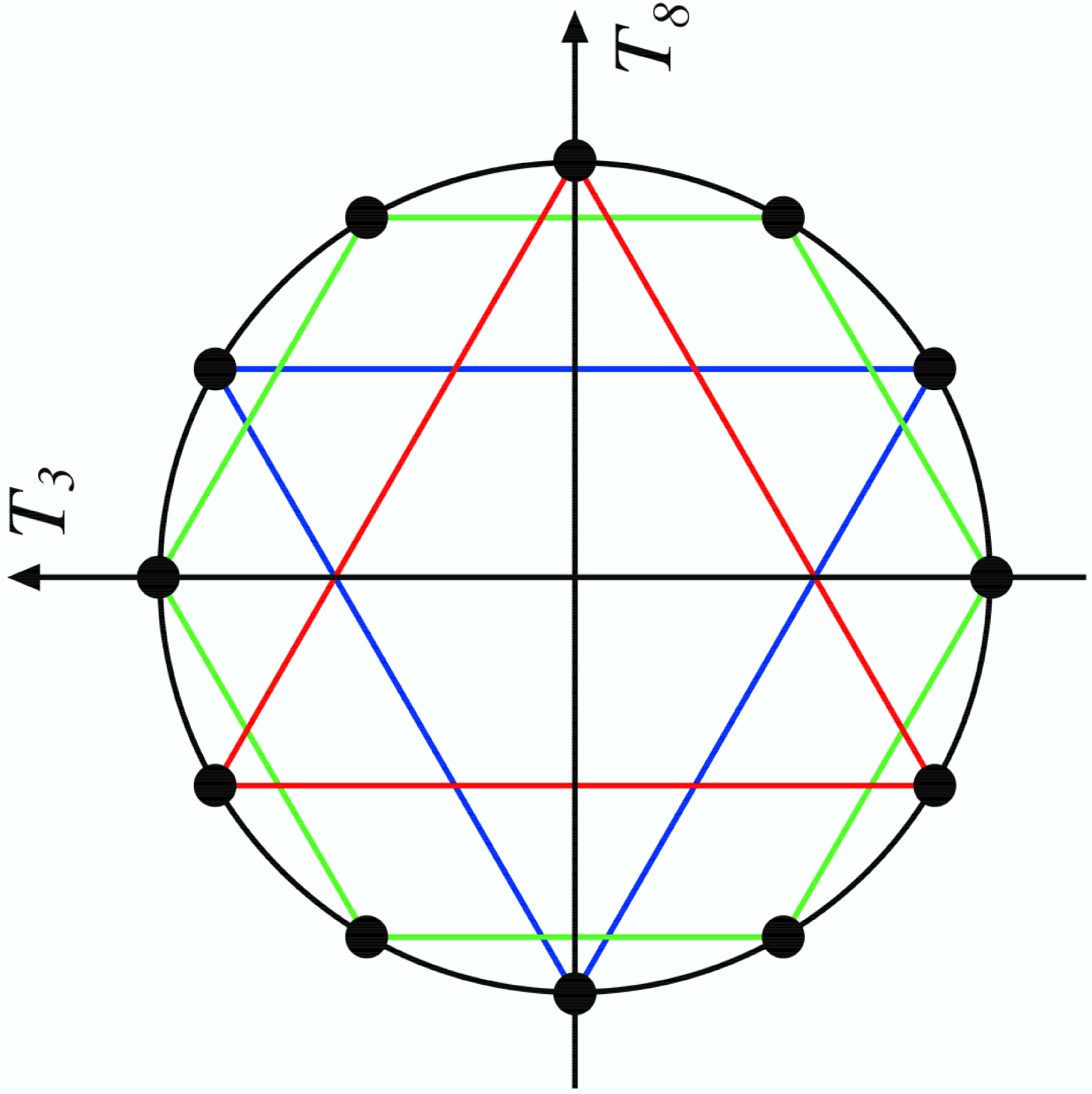}
\caption{
Orbital-dependent spectral functions $A_\alpha(\omega)=-\tfrac{1}{\pi}\text{Im}G_\alpha(\omega)$ for the equilibrium SOSM state corresponding to ${\mathcal T}_8>0$, ${\mathcal T}_3=0$ (left panel) and ${\mathcal T}_8<0$, ${\mathcal T}_3=0$ (middle panel). The spectra have been obtained for $U=1$ eV, $J=-0.25$ eV, $T=33$ K using the NCA impurity solver. Right panel: Equivalent SOSM states connected by permutations of the orbitals. The stable solutions are the three states equivalent to ${\mathcal T}_8>0$, ${\mathcal T}_3=0$ (red triangle). 
}
\label{figeq}
\end{figure}

\section{Results}

\subsection{Properties of the SOSM state}

We start by briefly discussing the essential properties of the SOSM states characterized by a coexistence of insulating and metallic orbitals \cite{footnote_ins}. In the model with density-density interactions ($\gamma=0$), the order is characterized by the quantities $\mathcal{T}_8$ and $\mathcal{T}_3$ defined as~\cite{Hoshino2016}
\begin{equation}
\mathcal{T}_8=(D_1+D_2-2D_3)/\sqrt{3}, \quad \mathcal{T}_3=D_1-D_2.
\end{equation}
Here, $D_\alpha=\langle n_{\alpha,\uparrow}n_{\alpha,\downarrow}\rangle$ is the double occupation in orbital $\alpha$.  
For $U/W=2.5$, $J=-U/4$ and $\beta W=140$ [$U=1$ eV, $J=-0.25$ eV, $T=33$ K], the orbital-dependent double occupations in the $\mathcal{T}_8>0$ state are
$D_1=D_2=0.465$, $D_3=0.167$, while in the $\mathcal{T}_8<0$ state, we find $D_1=D_2=0.303$, $D_3=0.479$ (NCA results). Figure~\ref{figeq} shows the corresponding orbital-dependent spectral functions. The $\mathcal{T}_8>0$ state (left panel) features two insulating and one metallic orbital. (Since $J<0$, the insulator is a paired Mott insulator, with an enhanced double occupation as compared to the metal, so that $D_1=D_2>D_3$.) In the  $\mathcal{T}_8<0$ state (middle panel), the first two orbitals are metallic and in the third orbital the electrons are paired, which results in $D_1=D_2<D_3$. The $\mathcal{T}_8=0$, $\mathcal{T}_3\ne 0$ state is difficult to stabilize in equilibrium. It has one insulating, one metallic and one bad-metallic orbital, whose spectral function exhibits a narrow gap in the quasi-particle peak. Here, the double occupations satisfy $D_1=D_3+\delta$ and $D_2=D_3-\delta$. By permuting the orbitals, we can generate three equivalent $\mathcal{T}_8>0$ and $\mathcal{T}_8<0$ states, and six equivalent $\mathcal{T}_3 \ne 0$ states, as illustrated in the right panel of Fig.~\ref{figeq}. All these states can be distinguished by their unique anisotropic conductivity. 

In Ref.~\onlinecite{Hoshino2016} it was shown that in equilibrium, the $\mathcal{T}_8>0$ state is the stable SOSM state, so that in the following, we will focus on the ultrafast switching between the different $\mathcal{T}_8>0$ states, i.e. between states with the following combination of metallic ($M$) and insulating ($I$) orbitals: $T_z\equiv (I,I,M)$, $T_y\equiv(I,M,I)$, and $T_x\equiv(M,I,I)$.

\begin{figure}[t]
\includegraphics[angle=-90, width=0.32\columnwidth]{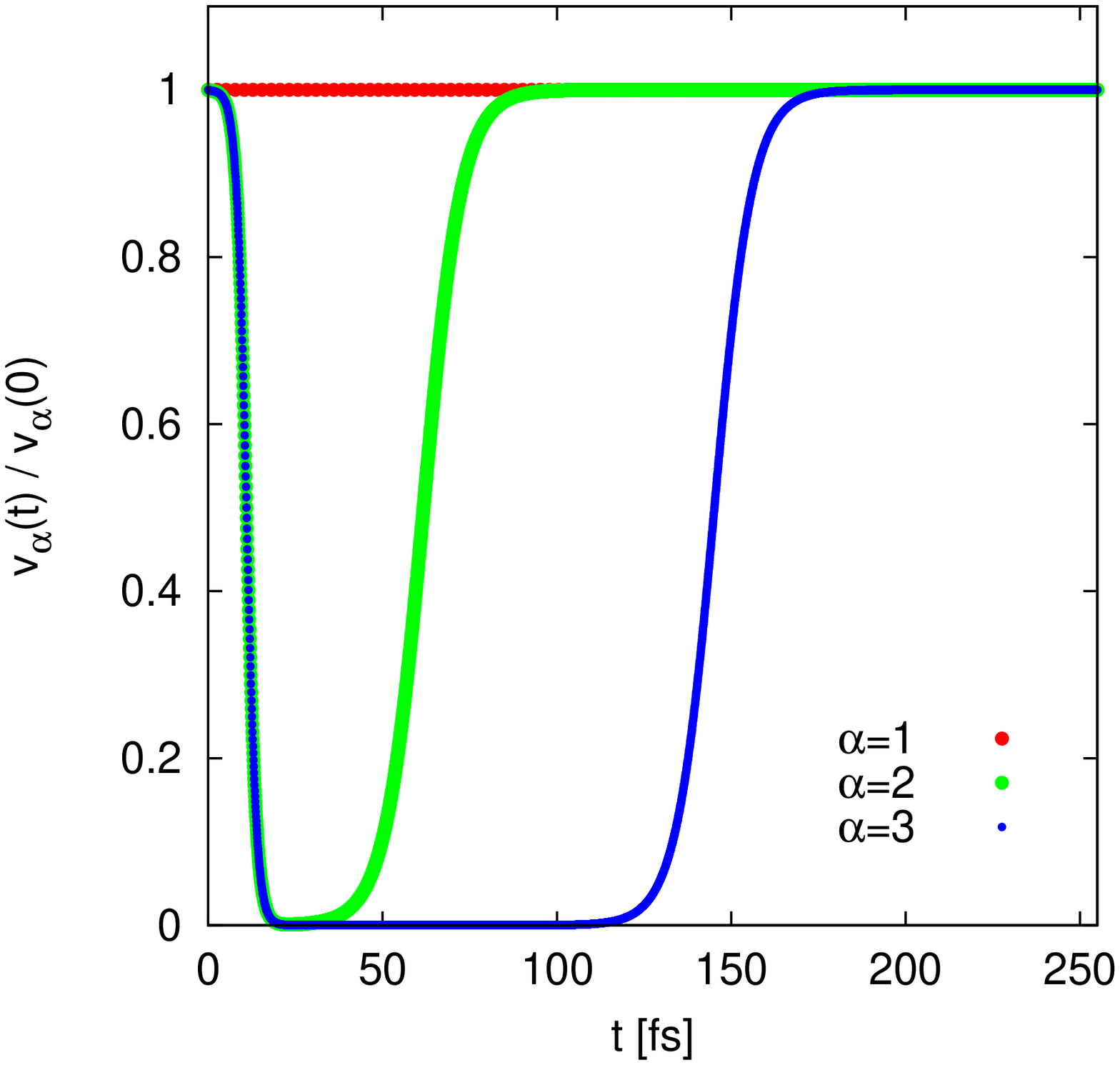}\hfill
\includegraphics[angle=-90, width=0.32\columnwidth]{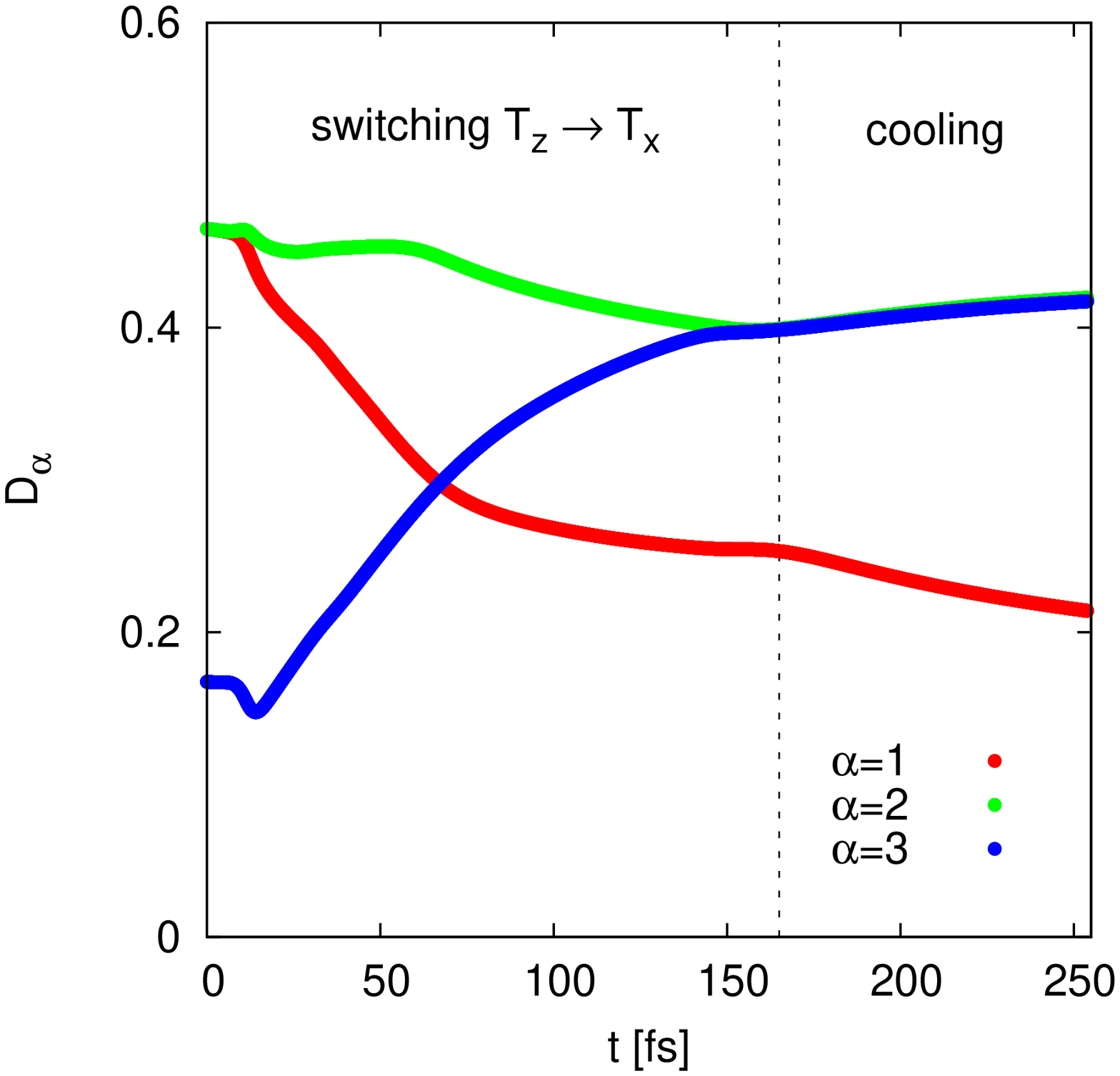}\hfill
\includegraphics[angle=-90, width=0.328\columnwidth]{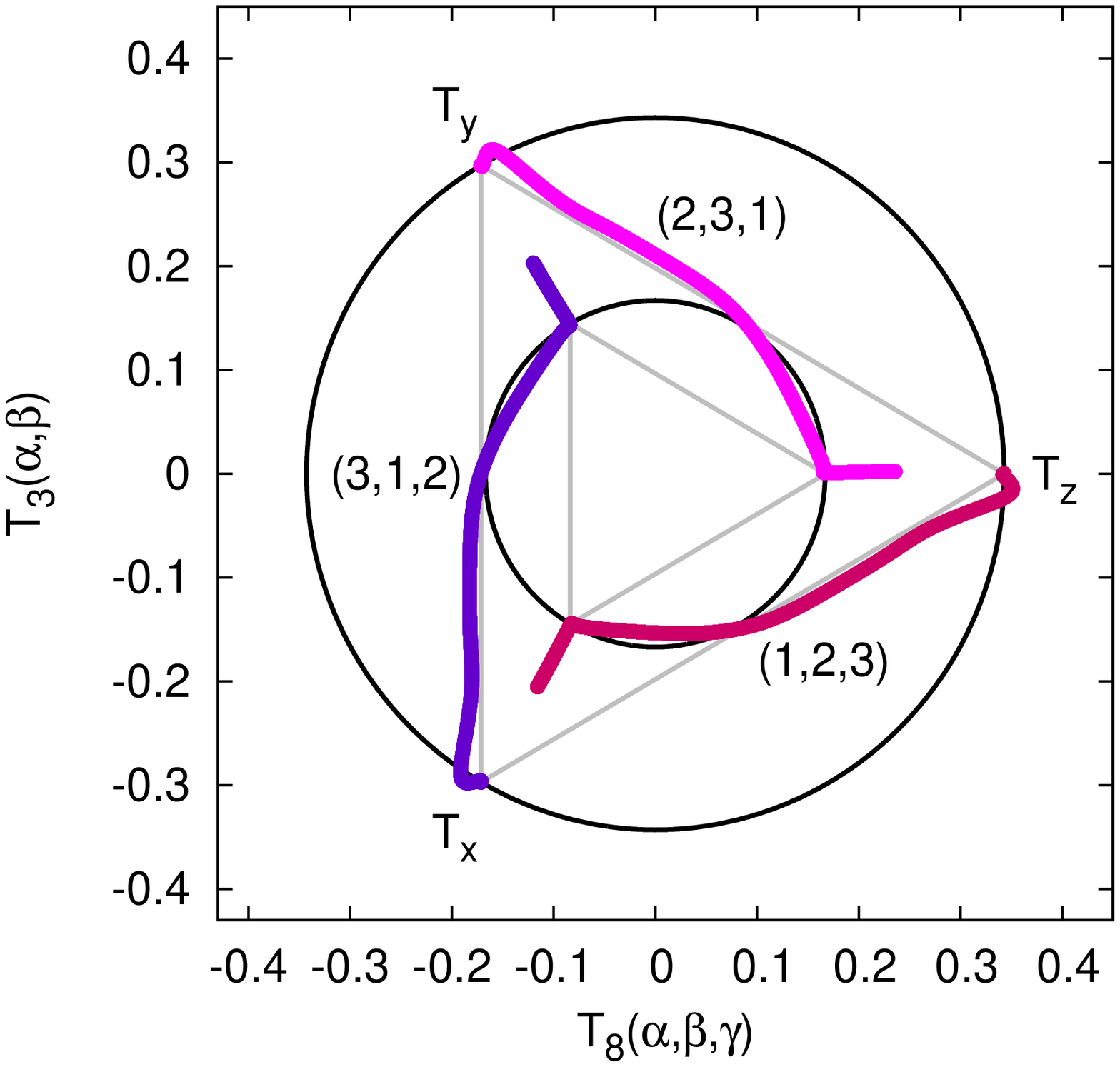}
\caption{
Switching between ${\mathcal T}_8>0$ states by suppression of hopping ($U=1$ eV, $J=-0.25$ eV, $T=33$ K). The order parameter can be defined in terms of the linear combinations ${\mathcal T}_8(\alpha,\beta,\gamma)=(D_\alpha+D_\beta-2D_\gamma)/\sqrt{3}$ and ${\mathcal T}_3(\alpha,\beta)=D_\alpha-D_\beta$. The initial state is $T_z=(I, I, M)$, with order parameter ${\mathcal T}_8(1,2,3)>0$, ${\mathcal T}_3(1,2,3)=0$, the state after the switching is $T_x=(M, I, I)$. The left panel shows the time evolution of the hopping parameters for the three orbitals, the middle panel the time evolution of the double occupancy, and the right panel the (clockwise) rotation of the order parameter in the ${\mathcal T}_8$-${\mathcal T}_3$ plane. 
}
\label{figswitch}
\end{figure}

\subsection{Switching between ${\mathcal T}_8>0$ states}

We found that switching between different SOSM states is possible by applying various types of time-dependent perturbations, such as (i) time-dependent crystal field splittings, (ii) time- and orbital-dependent interactions $U_\alpha$, and (iii) time-dependent electric fields. Protocols of the type (i) and, indirectly, (ii) may be realized by the selective driving of phonon modes, as discussed in Ref.~\onlinecite{Mitrano2015}. However, such a switching is limited by the intrinsic timescale of the phonons, and it involves (at least temporarily) the appearance of ordinary orbital moments. Here, we are interested in potentially faster switching protocols, which are limited only by some electronic timescale, and which do not induce any symmetry-breaking at the one-body level. A promising  pathway is the direct coupling of the electric field of the laser to the electrons. Static \cite{Freericks2008, Werner2015} or periodic \cite{Tsuji2011} electric fields are known to reduce the effective hopping in systems with well separated low-energy bands, as is the case in A$_3$C$_{60}$ \cite{Nomura2012}. For example, for an oscillating field $E(t) = E_0\cos(\Omega t)$ and in the limit of high-frequencies, the Peierls factor can be replaced by its time-average, given by $\mathcal{J}_{0}(eaE_0/\Omega)$, where $\mathcal{J}_0(x)$ is the Bessel function and $a$ the lattice spacing. To gain some insights into the switching dynamics, we will therefore first study simplified protocols, in which the {\em amplitude} of the hopping is modified in an orbital and time-dependent way, with the constraint that the orbital-dependent hoppings can only be reduced from their bare values, i.e., the term $v(t) e^{i\phi_\alpha(t)}$ in Eq.~(\ref{eq_field}) is replaced by a time-dependent real function $v_{\alpha}(t)\le v_\alpha(0)$. In section \ref{sec:field} we will then confirm that a switching can also be achieved by explicitly simulating the electric field, using similar protocols.

Figure~\ref{figswitch} illustrates a possible switching protocol. The left panel shows the time evolution of the hopping parameters, the middle panel the resulting evolution of the double occupation and the right panel the rotation of the order parameter in the ${\mathcal T}_8$-${\mathcal T}_3$ plane. Since we are manipulating the hopping, it is natural to think of the SOSM state as a state with orbital-dependent kinetic energy. The absolute value of the kinetic energy $K_\alpha$ is reduced for the insulating orbitals and enhanced for the metallic orbital. To switch from $T_z=(I,I,M)$ to $T_x=(M,I,I)$, we first suppress the hopping in orbitals $2$ and $3$. This breaks the symmetry between the insulating orbitals and leads to a reduction in the kinetic energy associated with the third orbital. After about 60 fs, the kinetic energies (and double occupations) in orbitals 1 and 3 reach similar values. If one would turn the hopping back on at this point, one would end up in a ${\mathcal T}_8<0$ state with two metallic and one insulating orbital. However, since this state is not the thermodynamically most stable one, we instead switch on the hopping in orbital $2$ only. This leads to an increase in $|K_2|$ and a faster evolution of orbitals $2$ and $3$ towards an insulating state with $K_2=K_3$ and $D_2=D_3$. At the same time, orbital $1$ turns metallic, so that we end up in the stable $T_x=(M,I,I)$ state. 

While especially the second part of the switching protocol leads to a reduction in the magnitude of the order parameter, the latter recovers to the thermal value as energy is dissipated to the bath, without adverse effect on the stability of the switched state. The cooling dynamics is evident in the evolution of the double occupation after time $t\approx 160$ fs and in the radial outward drift of the order parameter in the right panel. 

\begin{figure}[b]
\includegraphics[angle=-90, width=0.3\columnwidth]{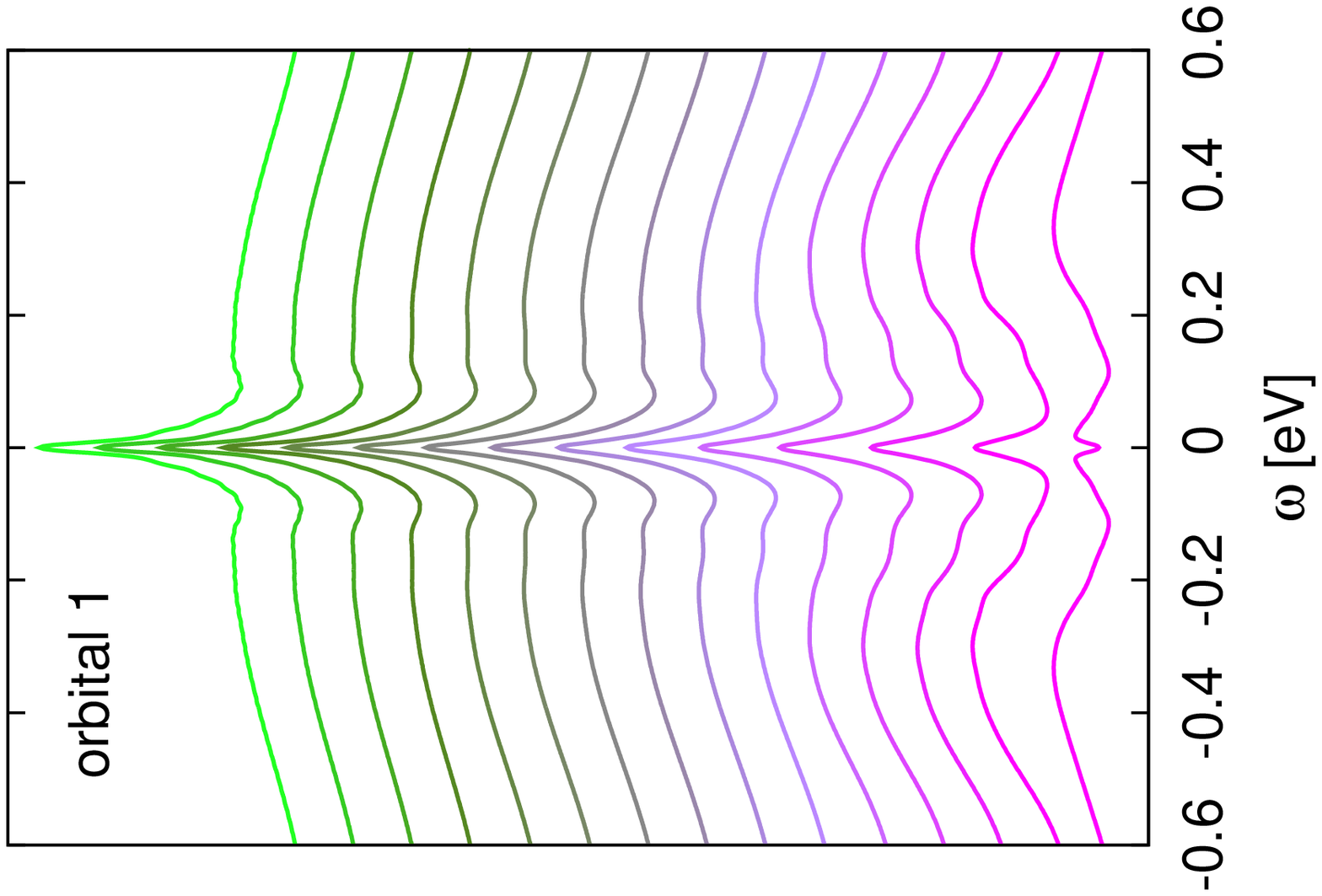}\hfill
\includegraphics[angle=-90, width=0.3\columnwidth]{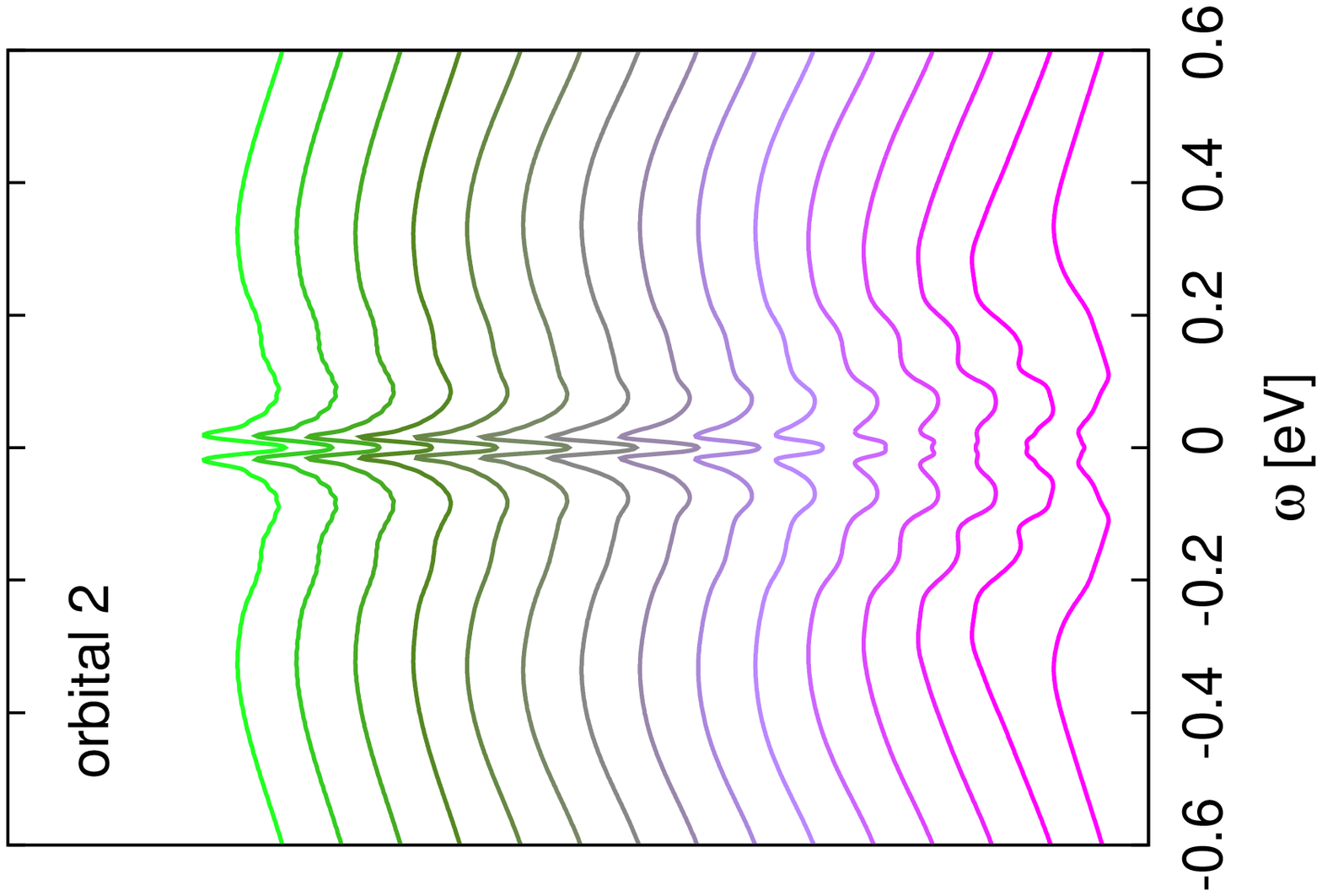}\hfill
\includegraphics[angle=-90, width=0.3\columnwidth]{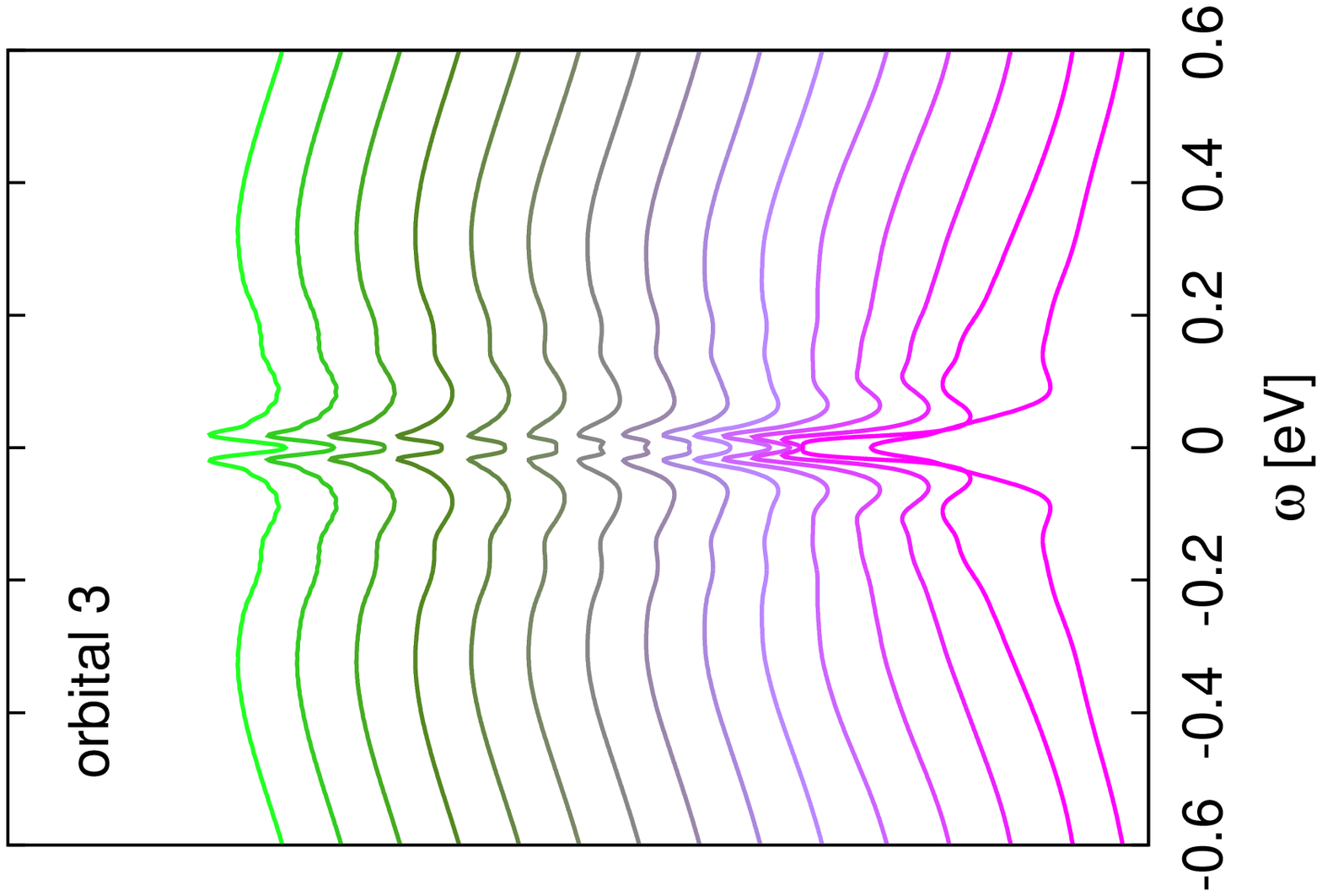}
\caption{
Switching from $T_z=(I,I,M)$ to $T_x=(M,I,I)$. From bottom to top: Local spectral functions for times $t=0, 14.5, 29, \ldots, 203$ fs. The spectra have been obtained by Fourier transformation on the time interval from $t$ to $t_\text{max}=280$ fs \cite{footnote_forward}.
}
\label{figspectra}
\end{figure}

To illustrate how the nature of the orbitals changes during the switching, we show in Fig.~\ref{figspectra} the time evolution of the orbital-resolved local spectral functions. The latter have been obtained as $A_\alpha(\omega,t)=-\frac{1}{\pi}\text{Im}\int_t^{t_\text{max}} dt' e^{i\omega(t'-t)}G^\text{ret}_\alpha(t',t)$, with $t_\text{max}=280$ fs \cite{footnote_forward}. 

The shortest switching time which can be realized depends on several factors, such as the minimal value of the hopping in the switching protocol, the maximum reduction in the order parameter that one is willing to tolerate, and the parameters of the heat bath. 
For an efficient cooling, the boson frequency $\omega_0$ has to be relatively small, and the boson coupling $g$ has to be sufficiently large. We choose $\omega_0=0.1$ and $g=1$ because these values allow to stabilize SOSM states with qualitatively correct features in the spectral functions, and an order parameter switching on numerically accessible timescales \cite{footnote_weakquench}.

\subsection{Effect of pair hopping}

\begin{figure}[b]
\includegraphics[angle=-90, width=0.32\columnwidth]{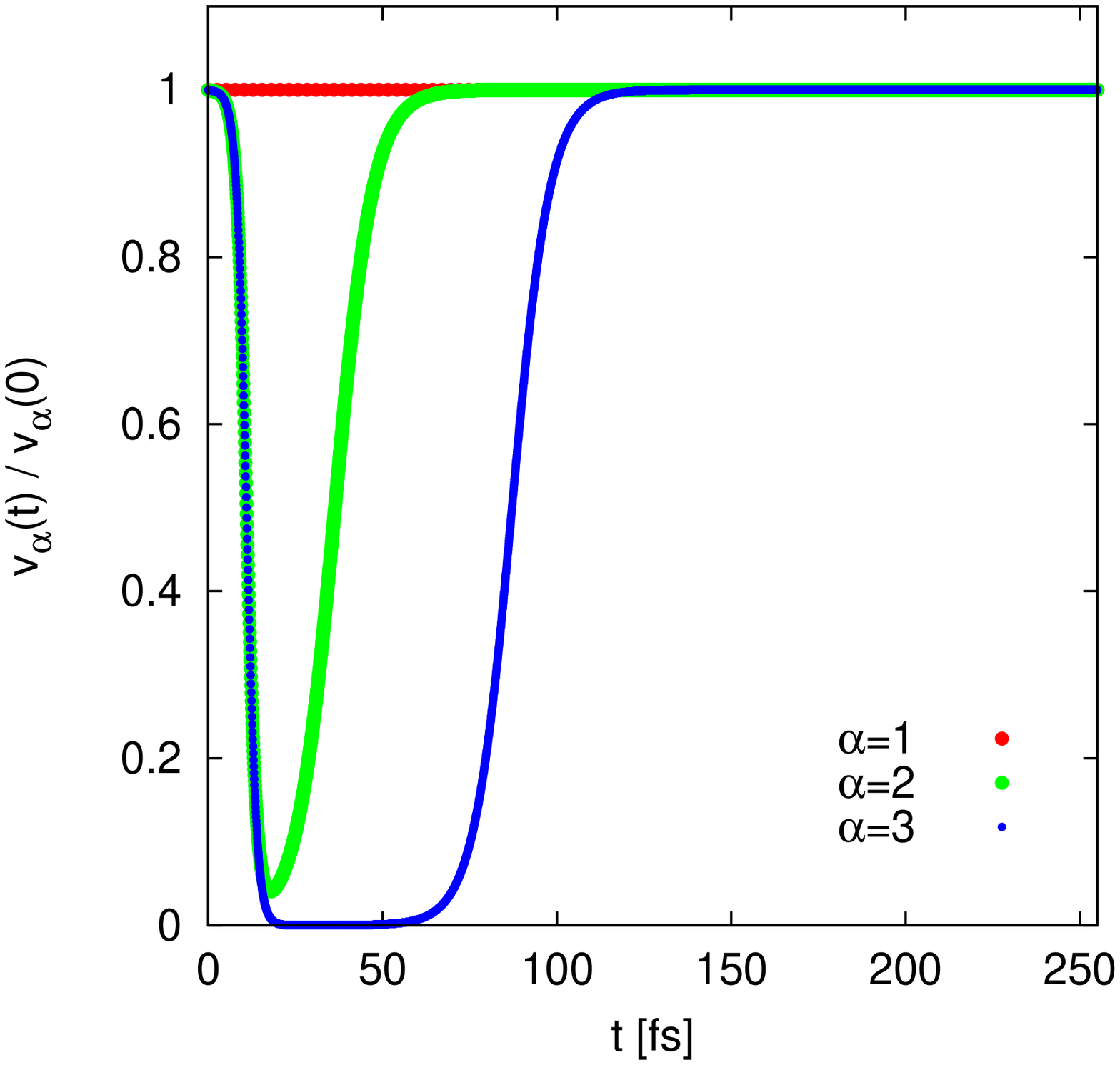}\hfill
\includegraphics[angle=-90, width=0.32\columnwidth]{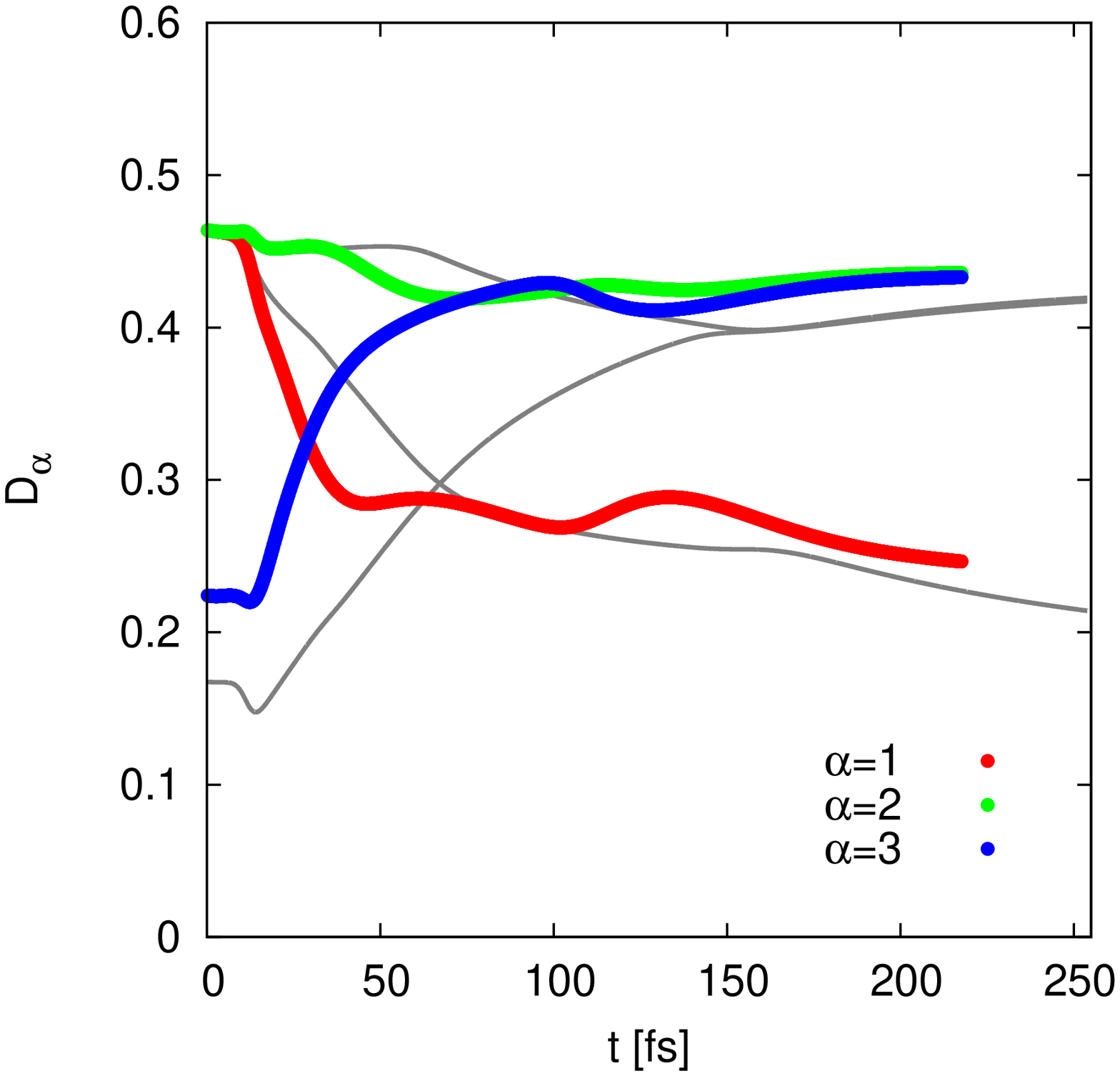}\hfill
\includegraphics[angle=-90, width=0.328\columnwidth]{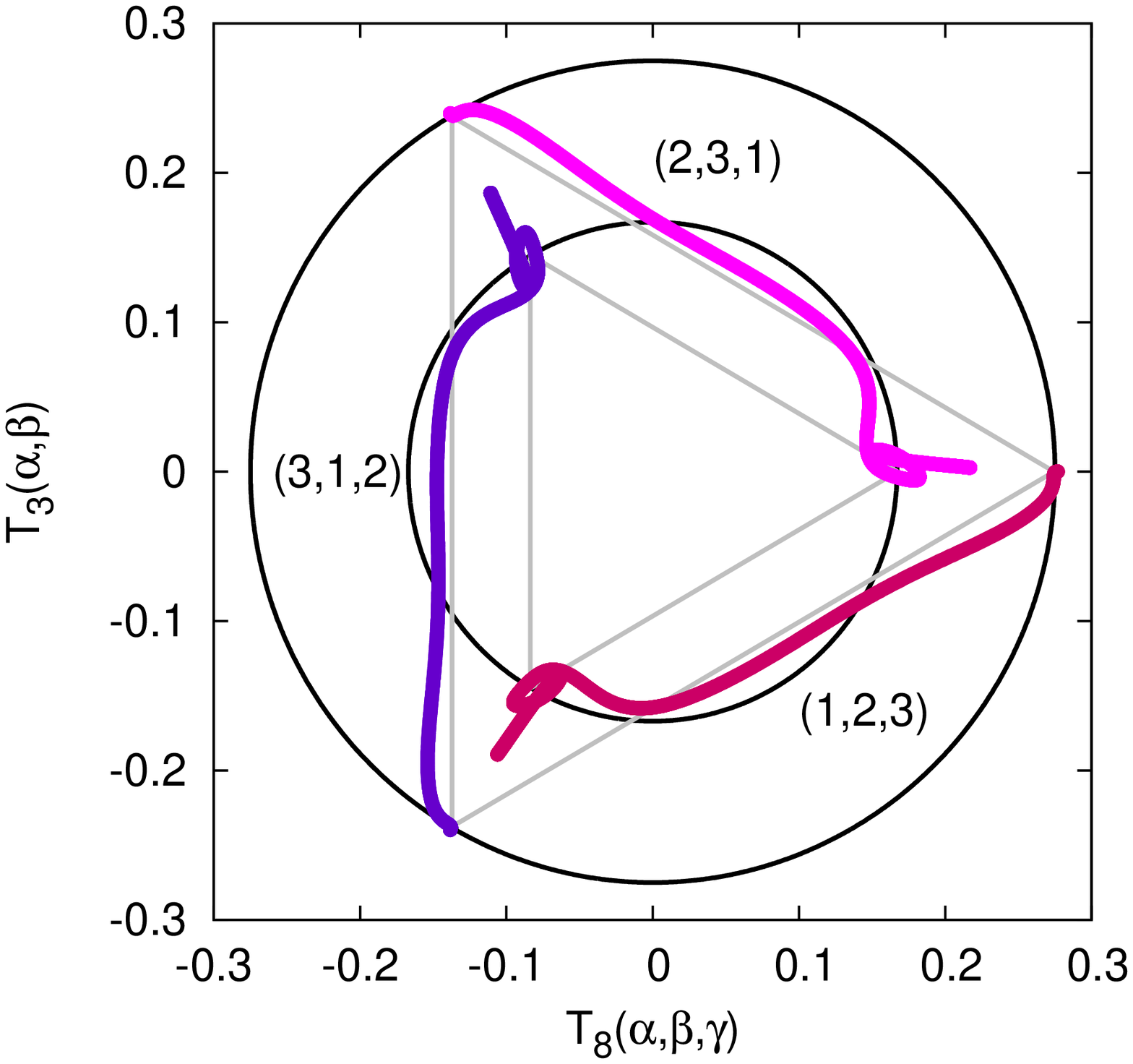}
\caption{
Switching between ${\mathcal T}_8>0$ states in the model with rotationally invariant interaction ($\gamma=1$).  
The left panel shows the switching protocol and the middle panel the time evolution of the double occupancy with gray lines taken from Fig.~\ref{figswitch} ($\gamma=0$) as a reference. The right panel shows the (clockwise) rotation of the order parameter in the ${\mathcal T}_8$-${\mathcal T}_3$ plane. 
}
\label{figswitch1}
\end{figure}

In this section we consider the model with rotationally invariant interaction ($\gamma=1$ in Eq.~(\ref{Hint})), but otherwise identical parameters. In this case, the system reacts more strongly to a reduction of the hopping, and the switching between different $\mathcal{T}_8>0$ states is faster than in the model with density-density interaction (Fig.~\ref{figswitch1}). Also, after a rapid switching, we observe small amplitude and phase oscillations of the order parameter around the new $\mathcal{T}_8>0$ state. 

\begin{figure}[t]
\includegraphics[angle=-90, width=0.32\columnwidth]{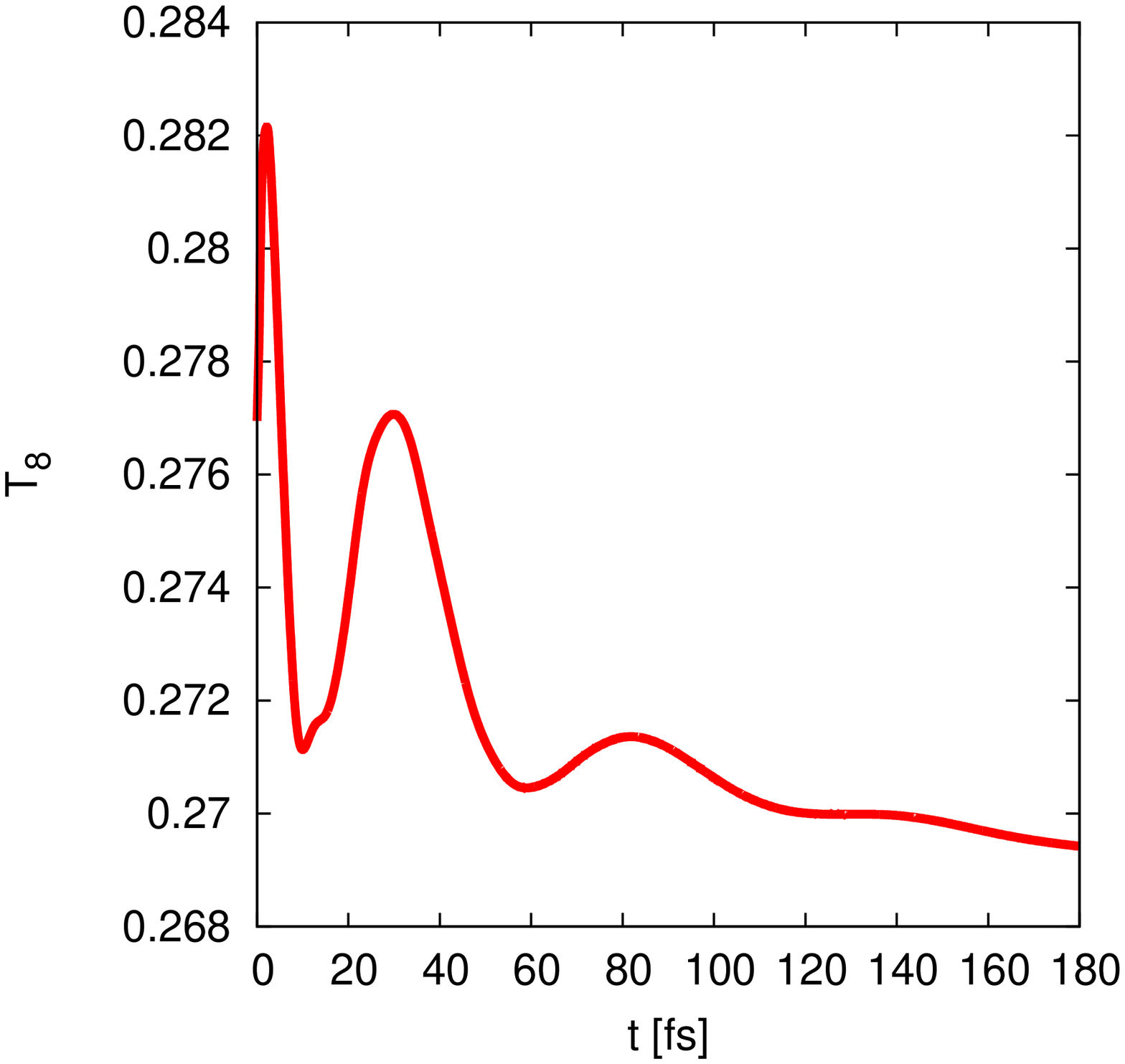}\hfill
\includegraphics[angle=-90, width=0.328\columnwidth]{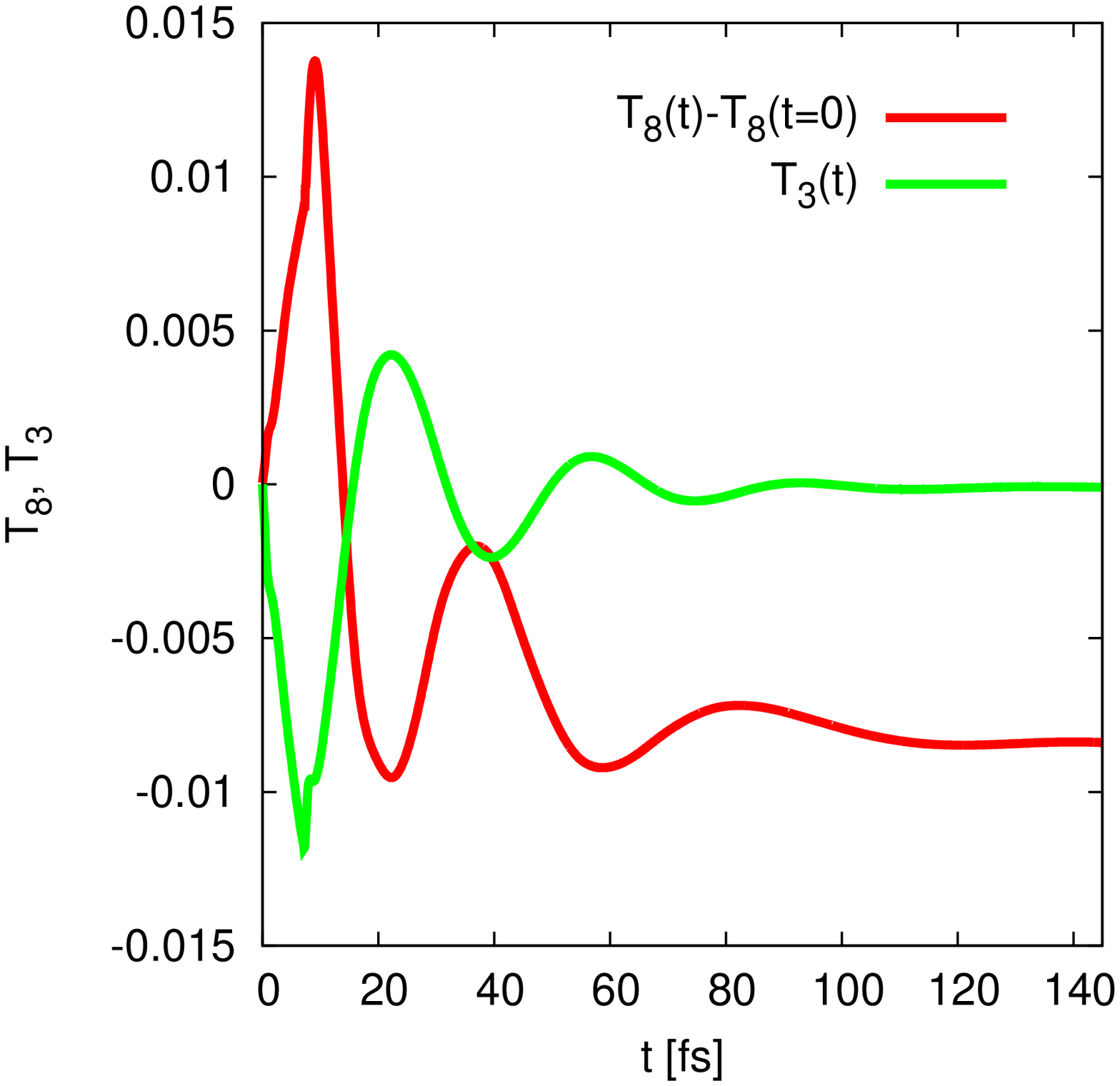}\hfill
\includegraphics[angle=-90, width=0.328\columnwidth]{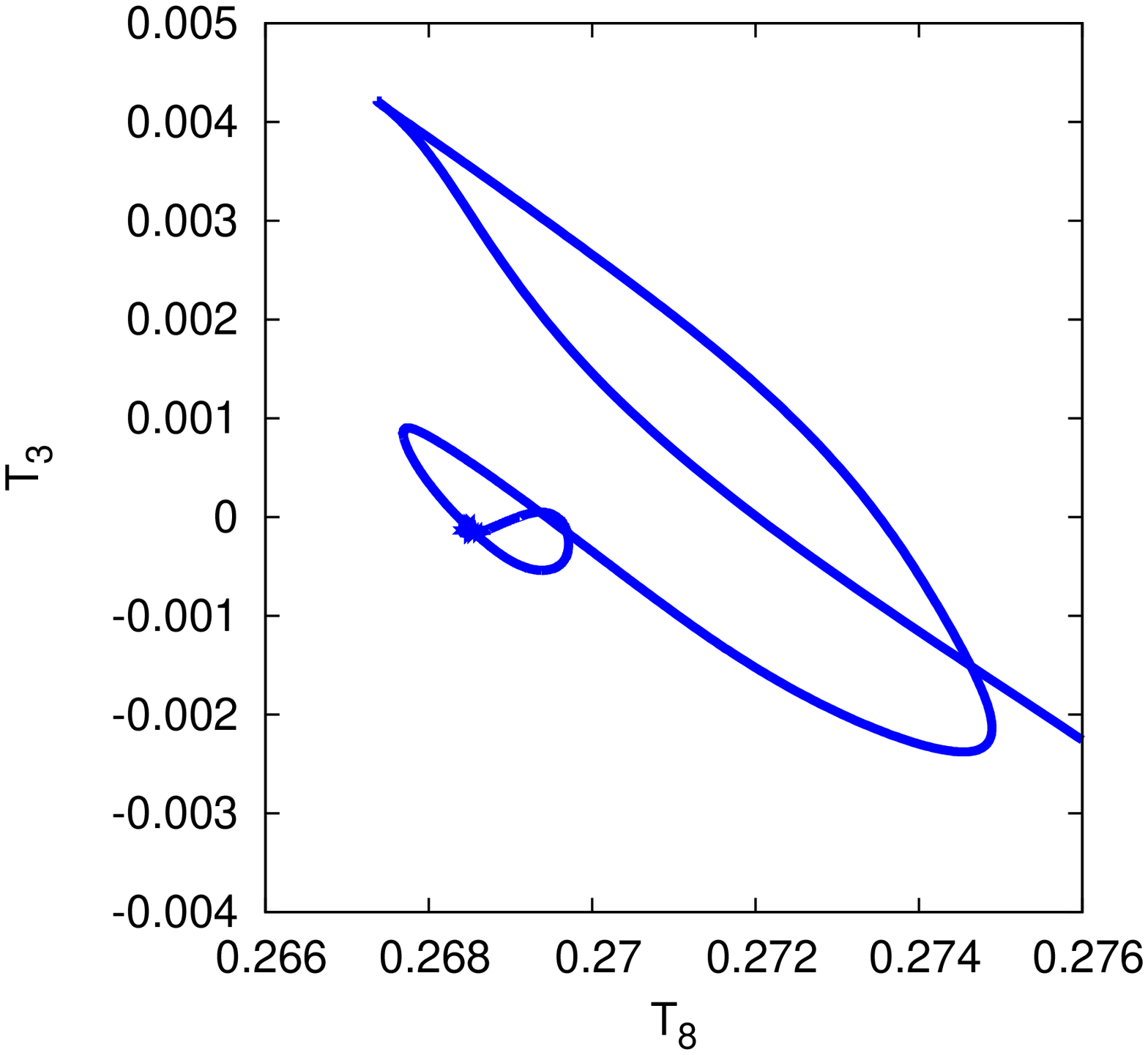}
\caption{
Excitation of amplitude and phase modes in the model with $\gamma=1$. The left panel shows ${\mathcal T}_8$ after a 20\% reduction in all hoppings at $t=0$. In the middle panel, the hopping quench is delayed by $\Delta t=7.3$ fs for orbitals 1 and 3, which allows to also excite oscillations in ${\mathcal T}_3$. The right panel shows the evolution in the ${\mathcal T}_8$-${\mathcal T}_3$ plane after this double-quench. 
}
\label{fighiggs}
\end{figure}

To excite the amplitude mode directly, we start in the $T_z$ state and quench the hopping for all orbitals to 80\% of the initial value at $t=0$. (In an experiment, this could be achieved by a strong field pulse with polarization aligned in the body diagonal of the crystal.) The resulting evolution of $\mathcal{T}_8$ is shown in the first panel of Fig.~\ref{fighiggs}. It exhibits well-defined, but strongly damped amplitude oscillations. Since this symmetric quench does not excite a phase mode, $\mathcal{T}_3$ remains zero during and after the pulse. To excite both modes, we apply an asymmetric quench, where the hopping in orbital 2 is reduced at $t=0$ and in orbitals 1 and 3 after a time delay of $\Delta t=7.3$ fs. The resulting dynamics of the $\mathcal{T}_8$ and $\mathcal{T}_3$ order parameters is shown in the middle panel. While the initial oscillation period of both modes is similar, it grows for subsequent oscillations in the case of the $\mathcal{T}_8$ mode, which indicates that the free energy minimum in the $\mathcal{T}_8$ direction becomes shallower due to heating, while it does not change substantially in the $\mathcal{T}_3$ direction. The third panel shows the evolution of the order parameter in the $\mathcal{T}_8$-$\mathcal{T}_3$ plane, which also illustrates that the period of the two modes drifts apart at later times. 

\subsection{Electric fields}
\label{sec:field}

In this final section, we show that switching is also possible by simulating the electric field of the laser. In the following, we will set $v_\alpha(t)=v$ to a constant and include a field through the Peierls phase in Eq.~\eqref{eq_field}. We considered two switching protocols: (i) switching through Wannier-Stark localization induced by static fields with $E>W$, and (ii) switching through a bandwidth renormalization induced by periodic fields with frequency $\Omega>W$. In case (i) we found that while it is possible to rotate the order parameter under the influence of quasi-static fields, it seems very difficult to switch off a quasi-static field in such a way that the order does not melt. In case (ii) we were able to rotate the order parameter without melting, although the heating effect is larger than in the case of a hopping quench. We consider in this section the model with density-density interactions, and the same parameters ($U=1$ eV, $J=-0.25$ eV, $T=33$ K) as in the previous simulations.  

\begin{figure}[t]
\includegraphics[angle=-90, width=0.32\columnwidth]{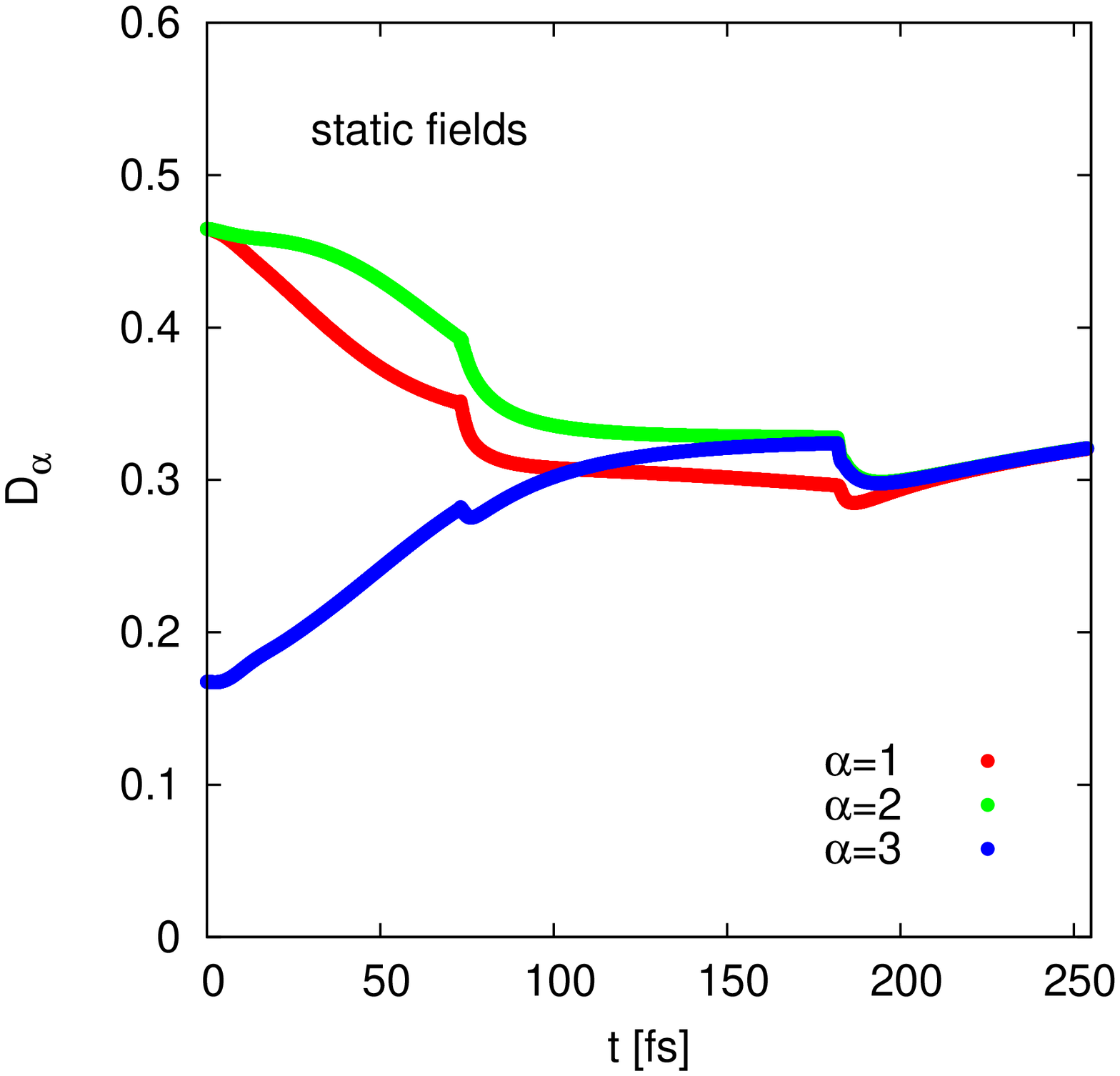}\hfill 
\includegraphics[angle=-90, width=0.328\columnwidth]{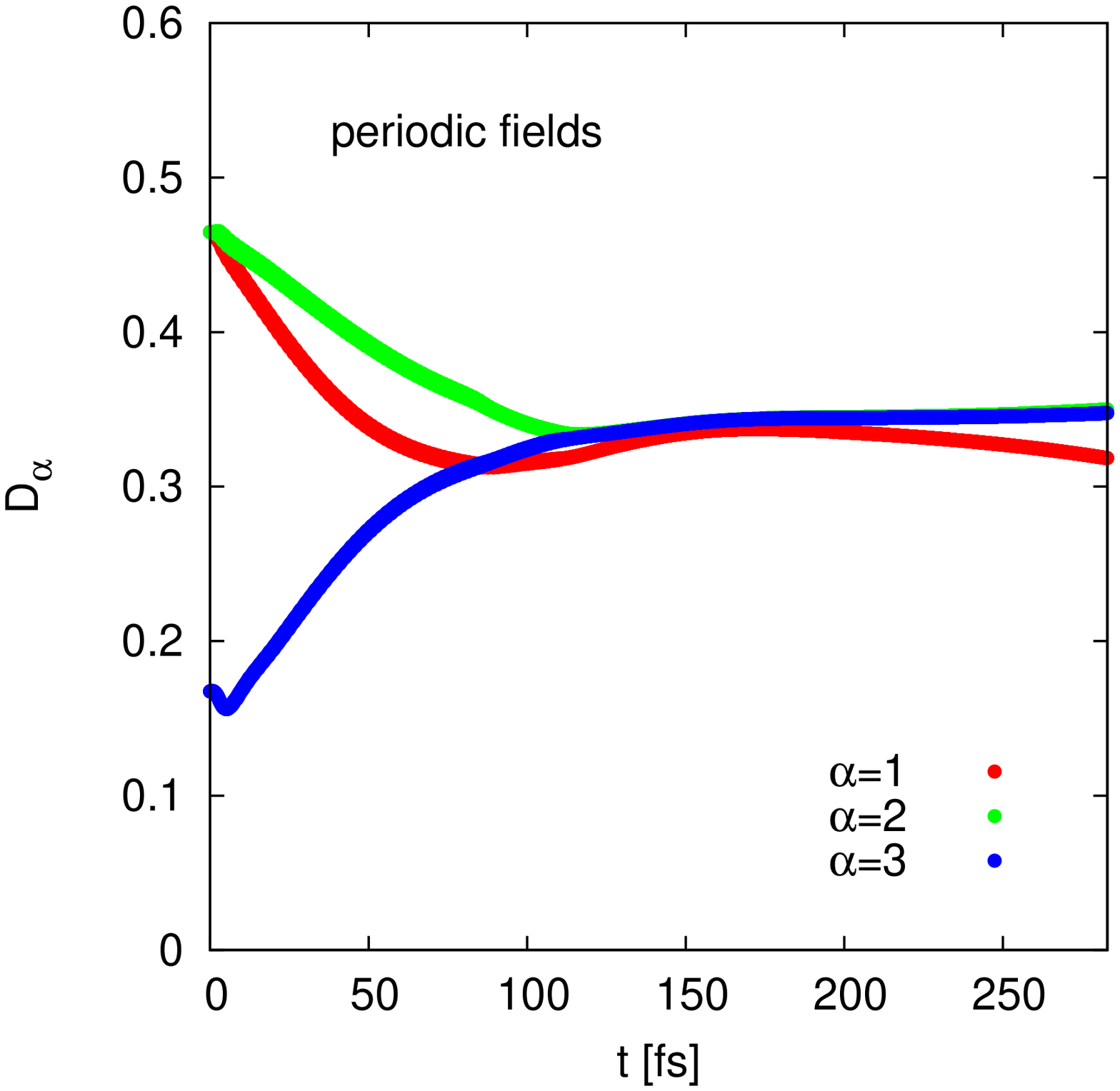}\hfill 
\includegraphics[angle=-90, width=0.328\columnwidth]{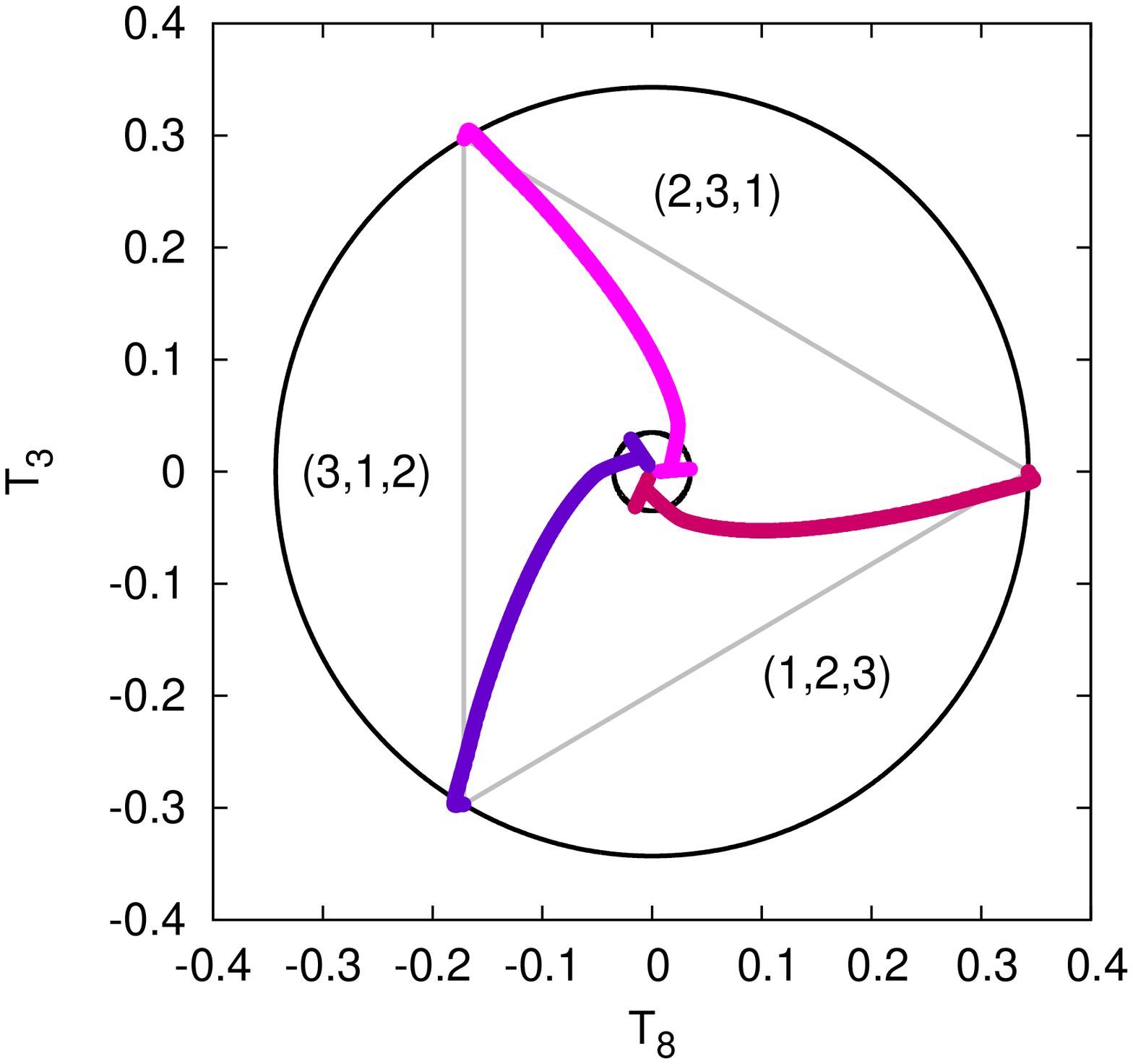}
\caption{
Switching by electric fields. Left panel: Switching by static fields with subsequent melting of the order during switch-off. Middle panel: Switching by periodic electric fields with amplitude $E_p=10$ and frequency $\Omega=3.33$ eV. The right panel shows the rotation in the ${\mathcal T}_8$-${\mathcal T}_3$ plane induced by the periodic fields.
}
\label{figper}
\end{figure}

In the quasi-static field switching illustrated in the left panel of Fig.~\ref{figper}, we apply fields of equal strength $E^{(2,3)}_s=8.33$ (measured in units of $W/ea$, with $a$ the lattice spacing \cite{footnote_fieldstrength}) to the orbitals 2 and 3 between time $t=73$ fs and $t=180$ fs, while orbital 1 is exposed to a small field of strength $E_s^{(1)}=0.11$ ($eaE_s^{(1)}$ equal to the initial gap size) from time $t=0$ to time $t=73$ fs. 

While this protocol succeeds in rotating the order parameter in a time of less than $190$ fs, both a rapid or smooth switch-off of the strong static field results in a melting of the order. The transient switching may however provide a seed for the symmetry breaking which occurs after cooling. It might also be possible to avoid the melting by applying a periodic modulation during the switch-off procedure, but in this case it is more natural to design a scheme which is based entirely on periodic fields. 

In the periodic driving protocol (middle panel of Fig.~\ref{figper}), we follow the same strategy as in the case of the hopping quench, and apply the field for $\Delta t=87$ fs in band 2 and for $\Delta t=115$ fs in band 3. The field amplitude is $E_p=10$ and the frequency $\Omega=3.33$ eV, which corresponds to an effective bandwidth renormalization by a factor $\mathcal{J}_0(eaE_p/\Omega)=0.67$ \cite{Tsuji2011}. A lower frequency or stronger effective band renormalization leads to stronger heating effects (note that the correlated density of states has a bandwidth of about 2 eV (Fig.~\ref{figeq})).

\section{Summary}

We studied the nonequilibrium dynamics of composite orbital order in a three band model with negative Hund coupling that captures the essential properties of fulleride compounds. The composite ordered state corresponds to the Jahn-Teller metal phase \cite{Zadik2015}, which exhibits a coexistence of two (paired) Mott insulating and one metallic orbital \cite{Hoshino2016}. In a cubic system with orbitals of $p_{x,y,z}$ type symmetry there exist three equivalent ordered states with metallic conduction in the $x,y,z$ direction, respectively, which may be used to store information. While the read-out of this information can be easily accomplished, a prerequisite for building a memory device based on these states is a reliable and fast switching mechanism between the three equivalent states. 

Using the nonequilibrium dynamical mean-field theory in combination with a strong-coupling (NCA) impurity solver we showed that the effective hopping reduction induced by strong quasi-static or periodic electric fields can induce the desired controlled rotation of the order parameter. While the switching speed depends on the energy dissipation rate and on the presence or absence of pair-hopping terms, we found that the write operation can in principle be accomplished in a time of O(100 fs), which is very fast compared to the O(100 ps) write speed of the currently fastest proposed technologies for non-volatile memories \cite{Vaskivskyi2016}. 
Furthermore, both the write and erase operations can be performed at the same speed, without relying on a thermal melting of the order. The field strengths and modulation speeds considered in this study are achievable with current day technology, and the $t_{1u}$ bands in fulleride compounds are separated from other bands by a gap of about 1 eV \cite{Nomura2012}. It therefore appears that the electric-field controlled switching of composite order in this class of materials on a sub-picosecond timescale is a realistic proposition, provided that suitable bulk crystals can be synthesized. Since the ordered states have no ordinary orbital moment, they should not induce static lattice distortions, so that the switching time is not limited by lattice dynamics.   

Our investigation pushes the limits of nonequilibrium DMFT simulations beyond the previously studied single band models and simple order parameters and demonstrates how this theoretical tool can contribute to the exploration of complex nonequilibrium phenomena in correlated multi-band systems.
While this work provides a proof of principle, a material-specific, realistic calculation would have to involve the hopping amplitudes for an fcc lattice, as well as a simulation of field-induced excitations into or out of higher lying bands. The latter effect is difficult to treat explicitly within nonequilibrium DMFT since the memory requirement grows rapidly with increasing number of bands, but it could be taken into account approximately by attaching a fermionic bath with a suitable density of states. We leave such a material-specific, quantitative simulation as an interesting topic for a future study.

\acknowledgements

The calculations were performed on the Beo04 cluster at the University of Fribourg. HS and PW acknowledge support from ERC starting grant No. 278023 and SH from JSPS KAKENHI Grant No. 16H04021. ME and PW thank the Kavli Institute for Theoretical Physics (National Science Foundation Grant No. NSF PHY-1125915) for its hospitality.


\begin{thebibliography}{99}
\bibitem{footnote_ins} There exist also completely insulating SOSM states, which we will however not consider in this work.
\bibitem{footnote_forward} Because of this forward time integration, the spectra at time $t=0$ are not identical to the equilibrium spectra (Fig.~\ref{figeq}), but affected by the subsequent nonequilibrium dynamics.
\end{thebibliography}
\end{document}